\documentclass[eqsecnum,twocolumn,showpacs,aps]{revtex4}

\usepackage{graphicx}
\usepackage{rotating}
\begin{document}

\title{Study of the  $pp \to pp\pi^+\pi^-$ Reaction in the Low-Energy Tail  of
  the Roper Resonance} 
\author{
J.~P\"atzold$^1$,
M.~Bashkanov$^1$,
R.~Bilger$^1$, 
W.~Brodowski$^1$,
H.~Cal\'en$^2$,
H.~Clement$^1$
C.~Ekstr\"om$^2$,
K.~Fransson$^3$,
J.~Greiff$^4$,
S.~H\"aggstr\"om$^3$,
B.~H\"oistad$^3$,
J.~Johanson$^3$,
A.~Johansson$^3$,
T.~Johansson$^3$,
K.~Kilian$^5$,
S.~Kullander$^3$,
A.~Kup\'s\'c$^2$,
P.~Marciniewski$^3$,
B.~Morosov$^6$,
W.~Oelert$^5$,
R.J.M.Y.~Ruber$^3$,
M.~Schepkin$^7$,
W.~Scobel$^4$,
J.~Stepaniak$^8$,
A.~Sukhanov$^6$,
A.~Turowiecki$^9$,
G.J.~Wagner$^1$,
Z.~Wilhelmi$^9$,
J.~Zabierowski$^{10}$,
J.~Zlomanczuk$^3$
}
\affiliation{
$^1$Physikalisches Institut der Universit\"at T\"ubingen, 
Morgenstelle 14, D-72076  T\"ubingen, Germany \\
$^2$The Svedberg Laboratory, S-751 21 Uppsala, Sweden \\
$^3$Department of Radiation Sciences, Uppsala University, S-751 21 Uppsala,
Sweden \\
$^4$Institut f\"ur Experimentalphysik der Universit\"at Hamburg, Germany \\
$^5$IKP - Forschungszentrum J\"ulich GmbH, D-52425 J\"ulich, Germany \\
$^6$Joint Institute for Nuclear Research Dubna, 101000 Moscow, Russia \\
$^7$Institute for Theoretical and Experimental Physics, 117218 Moscow, Russia
\\ 
$^8$Soltan Institute for Nuclear Studies, PL-00681 Warsaw, Poland \\
$^9$Institute of Experimental Physics, Warsaw University, PL-0061 Warsaw,
  Poland \\
$^{10}$Soltan Institute for Nuclear Studies, PL-90137 L\'odz, Poland
}
\date{\today}

\begin{abstract}
Exclusive measurements of the $pp \to pp\pi^+\pi^-$ reaction have been carried
out at $T_p = 775$ MeV at CELSIUS using the PROMICE/WASA setup. Together with
data obtained at lower energy they point to a dominance of the Roper excitation
in this process. From  the observed interference of its decay routes $N^\ast
\to N\sigma$ and $N^\ast \to \Delta \pi \to N\sigma$ their energy-dependent
relative branching ratio is determined.
\end{abstract}

\pacs{13.75.-n, 14.20.Gk, 25.40.Ve}

\maketitle

\def\lapprox{{\raise0.5ex\hbox{$<$}\hskip-0.80em\lower0.5ex\hbox{$\sim$}
}}
\def\gapprox{{\raise0.5ex\hbox{$>$}\hskip-0.80em\lower0.5ex\hbox{$\sim$}
}}

The Roper resonance $N^\ast(1440)$ with $I(J^P) = 1/2(1/2^+)$ is presently
known as the second excited state of the nucleon \cite{pdg00}. In contrast to
the first excited state, the $\Delta(1232)$, and also higher-lying resonances
the $N^\ast(1440)$ is still poorly understood both theoretically and
experimentally. Since it is hardly excited by electromagnetic probes
and has the same quantum numbers as the nucleon, it has been
interpreted  as the breathing mode monopole excitation 
of the nucleon. A recent theoretical work \cite{kre00,her02} finds the Roper
excitation to rest solely on meson-nucleon dynamics, whereas another recent
investigation \cite{mor99} proposes it to be actually two resonances with one
being the breathing mode and the other one a $\Delta$ excitation built on top
of the $\Delta(1232)$. In all these aspects the decay modes of the Roper
resonance into the $N\pi\pi$
channel play a crucial role. The simplest decay is $N^\ast \to
N(\pi\pi)_{I=l=0} := N\sigma$, i.e., the decay into the $\sigma$ channel. A
competitive and according to present knowledge \cite{pdg00} actually much
stronger decay channel is the sequential Roper decay via the $\Delta(1232)$
resonance,
$N^\ast \to \Delta\pi$. However, this decay channel is not very well defined,
in particular not orthogonal to the $N\sigma$ channel, since the $\Delta$ is
also unstable and decays nearly as fast as the Roper does. In fact, most of
this decay will end up again in the $N\sigma$ channel and thus will interfere
with the direct $N^\ast \to N\sigma$ decay.

In a previous work, the first exclusive measurement of the $pp \to
pp\pi^+\pi^-$ reaction at $T_p = 750$ MeV \cite{bro02}, we have shown that
at energies not far above threshold  this reaction can be well described by
 dominant 
$\sigma$ exchange in the initial $NN$ collision with subsequent excitation of
the Roper resonance in one of the nucleons. This result, which is in agreement
with theoretical predictions of the Valencia group \cite{alv98}, exhibits this
reaction to be unique in the sense that it selectively provides the excitation
mode $``\sigma$'' N $\to N^\ast$ (where ``$\sigma$'' stands for the $\sigma$
exchange), which is not accessible in any other basic reaction process leading
to the Roper excitation.

In this work we present new data from an exclusive measurement of the $pp \to
pp\pi^+\pi^-$ reaction at $T_p = 775$ MeV. Together with the data at $T_p =
750$ MeV they are analysed with particular emphasis on the interference of the
decay routes $N^\ast \to N\sigma$ and $N^\ast \to \Delta \pi \to N\sigma$. The
data have been taken at the CELSIUS storage ring using the PROMICE/WASA
detector with a cluster jet H$_2$ target \cite{cal96}. Protons and $\pi^+$
particles have been registered in the forward detector part covering the polar
angles $4^\circ \leq \Theta_{Lab} \leq 21^\circ$. The particles have been
identified by the $\Delta E-E$ method, the stopped $\pi^+$ particles in
addition by their delayed pulse from subsequent muon decay. From the measured
four-momenta of the two registered

\begin{figure}
\resizebox{0.45\textwidth}{!}{%
\includegraphics{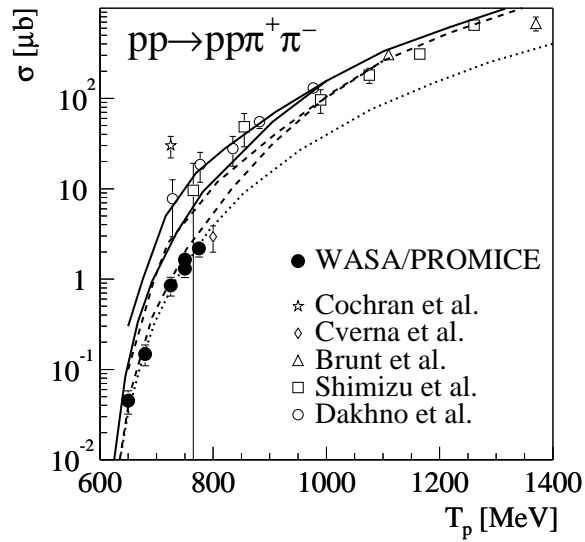}
}
\caption{Energy excitation function of the integral cross section of the $pp
  \to pp\pi^+\pi^-$ reaction. The WASA/PROMICE data (black points) from this
  work and Ref. [5] are compared with previous data (open symbols) [10-15] and
  predictions [6] with two different parameter sets (solid and dashed line),
  with and without $pp$ final state interaction (upper and lower lines
  respectively) and a phase space distribution adjusted to our data (dotted
  line).}
\end{figure}

\noindent protons and the identified $\pi^+$ the full
$pp\pi^+\pi^-$ events have been reconstructed by kinematical fits with one
overconstraint. Detector efficiencies and acceptance have been obtained from
Monte-Carlo (MC) simulations of the detector response \cite{pae02}. The
absolute normalization of the data has been obtained by monitoring the
luminosity of the experiment by the simultaneous measurement of the elastic
scattering and its comparison to data from literature \cite{arn97}. The total
cross section for the $pp \to pp\pi^+\pi^-$ reaction obtained for $T_p = 775$
MeV is $\sigma_{tot} = 2.2(5) 
\mu$b and shown in Fig. 1, together with previous results
\cite{bro02,cve81,coc72,shi82,dak83,bru69,joh00}. Our value is an order of
magnitude 
below the bubble chamber results \cite{dak83}, in agreement with our findings
at lower energies \cite{bro02}. The estimated uncertainty of about 20\% is due
to \cite{pae02} statistical uncertainties 
\noindent in the collected $pp$ elastic (1\%)
and $pp\pi^+\pi^-$ (5\%) events as well as systematic uncertainties in the
selection 
of $pp$ elastic (8\%) and $pp\pi^+\pi^-$ (12\%) events, uncertainty in the
lifetime of the data acquisition system (8\%) and uncertainties in the
extrapolation to full solid angle (6\%). For $T_p = 750$ MeV we show two values
for $\sigma_{tot}$. The upper one is the previously published value
\cite{bro02}, the lower one has been derived from a subsample of those data
using the same  event selection criteria as applied now for the 775
MeV data.  In order to test the robustness of the analysis the event selection
criteria have been slightly modified compared to the ones used
previously \cite{bro02}. However, within uncertainties both values agree with
each other. 

\begin{figure}
\resizebox{0.5\textwidth}{!}{%
\includegraphics{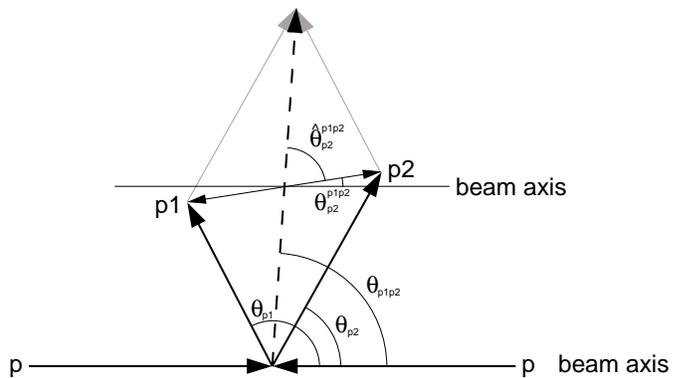}
}
\caption{Definition of the different scattering angles in the subsystem of
  particles, here for the case of two particles $(p1,p2)$ resulting from the
  reaction in the overall centre of mass system. For simplicity, the
  figure shows a non-relativistic construction. The other particles are not
  shown. For details see text.}
\end{figure}

Differential cross sections are shown in Figs. 3-5, which will be
discussed in the following.
As pointed out in Ref. \cite{bro02}, the  proton angular distribution in the
overall center of mass system (cms) is governed by the meson exchange between
the colliding protons. In particular, it should be described by
$\sigma$-exchange leading to $\sigma(\Theta_p) \sim 1 - a \cos^2(\Theta_p)$
with $a > 0$ given by the amplitude for $\sigma$ exchange while $a < 0$ would
be typical for $\pi$-exchange. The data at $T_p =
750$ MeV were well described by this ansatz. This holds also for the new data
at $T_p = 775$ MeV \cite{pae02}. Instead of looking at the angular
distribution in the overall cms it appears more instructive to look at the
angular distributions in the $pp$ subsystem. The definition of angles is
illustrated in Fig. 2. Let us denote the scattering angles of particles 1 and
2 in the overall  cms by $\Theta_{p_1}$ and $\Theta_{p_2}$, respectively, and
the scattering angle of the center of mass motion of both protons (summed
momenta)  in the overall cms by $\Theta_{p_1p_2}$. Within the rest frame of
the two particles ($p_1p_2$ subsystem) then two angles can be defined, the
scattering angle of $p_2$ either with respect to the beam axis,
$\Theta^{p_1p_2}_{p_2}$, or with respect to the summed momenta of $p_1$ and
$p_2$ in the overall cms. The latter  angle is denoted by
$\widehat{\Theta}^{p_1p_2}_{p_2}$. Note that due to the indistinguishability
of the two protons all these angular distributions have to be symmetric about
$90^\circ$. Since in this reaction the two protons are emitted dominantly
back-to-back in the overall cms (see distributions of the opening angle
$\delta_{pp}$ between the two protons in Refs. \cite{bro02,pae02}) the
$\Theta^{pp}_p$ distribution (Fig. 3) is very close to the $\Theta_p$
distribution and

\begin{figure}
\resizebox{0.5\textwidth}{!}{%
\includegraphics{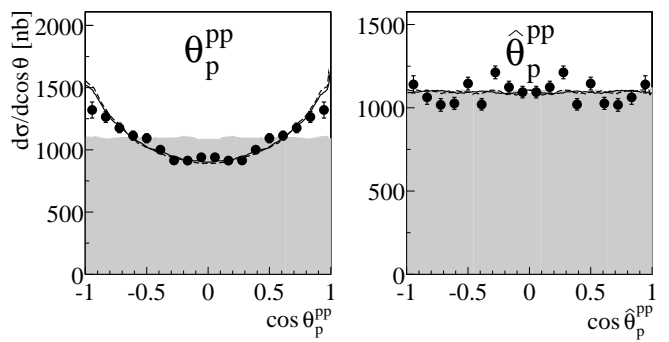}
}
\caption{Proton angular distributions in the $pp$ subsystem at $T_p = 775$
  MeV. In the left diagram the proton emission angle $\Theta^{pp}_p$ is taken
  relative to the protons' summary momentum in the overall cms. Shaded areas
  give the phase space distribution, whereas dashed and solid lines show MC
  simulations according to ansatz (1) and (2), respectively.}
\end{figure}

\noindent  
hence also exhibits a $(1 - a \cos^2 \Theta^{pp}_p)$
dependence. Shown in Fig.~3 is in addition the $\widehat{\Theta}^{pp}_p$
distribution. This distribution reflects the scattering situation of the two
protons within their subsystem. The observed distribution is isotropic, i.e. we
indeed see the outgoing protons to be in relative s-wave, as anticipated in
Ref.~\cite{bro02}.

We note in passing that also the pion angular distribution in the overall cms
\cite{pae02} is flat as well --- as was the case at $T_p = 750$ MeV,
too. Since here only 
the process $NN \to \Delta\Delta \to NN\pi\pi$ leads to nonisotropic
angular distributions \cite{bro02,alv98,pae02}, we conclude that this process,
which is expected to contribute substantially at energies $T_p > 1000$ MeV
\cite{alv98} is not yet of relevance at $T_p = 775$ MeV.

To see whether the
reaction proceeds via $N^\ast$ excitation, we inspect the measured
distribution of the $p\pi^+\pi^-$ invariant mass $M_{p\pi^+\pi^-}$
(Fig. 4). Compared to phase space the data are substantially enhanced near
the high-energy end, compatible with the low-energy tail of the $N^\ast$
excitation and reproduced by the appropriate calculations for $N^\ast$
excitation, as will be discussed in detail below.
As in Ref. \cite{bro02} we conclude that the process ``$\sigma$''$N \to
N^\ast$ is 
indeed the one which drives the reaction $pp \to pp\pi^+\pi^-$ at the energies
considered here. 

\begin{figure}
\resizebox{0.45\textwidth}{!}{%
\includegraphics{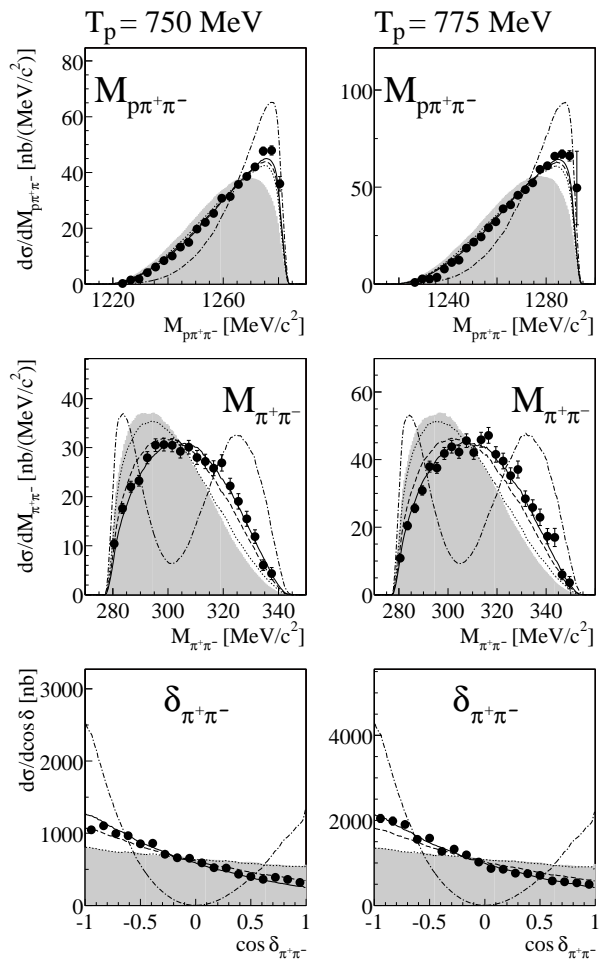}
}
\caption{Invariant masses $M_{p\pi^+\pi^-}$ and $M_{\pi^+\pi^-}$ as well as
  the opening angle $\delta_{\pi^+\pi^-}$ between the two pions for both 750
  (left) and 775 MeV (right) beam energies. The data are shown in comparison
  to  phase space (shaded area) and MC simulations for pure decays $N^\ast \to
  N\sigma$ (dotted), $N^\ast \to \Delta\pi$ (dashed dotted) and their
  interference with $c' = -37$ (dashed),  and --61 (solid)
  (GeV/c)$^{-2}$ using eq. (2).}
\end{figure}

We now examine the $N^\ast$ decay process as exhibited by the
data. In the analysis of the 750 MeV data two versions for the $N^\ast$
decay amplitude have been proposed \cite{bro02}:
$${\cal A} \sim 1 + c~{\bf k}_1 \cdot {\bf k}_2 (3 D_{\Delta^{++}} +
D_{\Delta^0}) \eqno(1)$$
and
$${\cal A} \sim (1 + c'~{\bf k}_1 \cdot {\bf k}_2)D_{\Delta^{++}}~. \eqno(2)$$
In the full reaction amplitude this factor ${\cal A}$ complements the
propagators for $\sigma$ 
exchange and $N^\ast$ excitation as well as the expression describing the
final state interaction between the two outgoing protons in relative
$s$-wave. Here $D_{\Delta^{++}} = 1/(M_{p\pi^+} - M_{\Delta^{++}} +
\frac{i}{2} \Gamma_{\Delta^{++}})$ and $D_{\Delta^0}$ defined analogously, are
the $\Delta$ propagators. The constant 1 stands for the process $N^\ast \to
N\sigma$ and the scalar product ${\bf k}_1 \cdot {\bf
  k}_2$ of the pion momenta ${\bf k}_1$ and ${\bf k}_2$ for the double
$p$-wave decay of the route $N^\ast \to \Delta \pi \to N\sigma$. 
For simplicity we have neglected the spinflip term $i {\bf s} \cdot ({\bf k}_1
\times {\bf k}_2)(3 D_{\Delta^{++}} - D_{\Delta^0})$
in this decay channel  with ${\bf s}$ being the
nucleon spin. This term describes the transition $N^\ast \to \Delta\pi \to
N(\pi\pi)_{I=l=1}$. It does not interfere with the other terms, is smaller
than the ${\bf k}_1 \cdot {\bf k}_2$ term by a factor of 16 in the cross
section and hence does not influence significantly the conclusions in this
paper.  Whereas in
Refs. \cite{bro02,alv98} this scalar product has been calculated in the overall
cms, we  here use  the more appropriate $N^\ast$ system. We note, however,
that the difference is tiny, since near threshold the static limit
approximation is reasonably valid. 

Ansatz (1) represents the leading term of the two-pion decay of the Roper
resonance as worked out by the Valencia group \cite{alv98}. 
The constant $c$ (and correspondingly $c'$ in ansatz (2)) gives the relative
strength between the two 
decay routes and is treated in the following as the parameter to be adjusted to
the data, which will enable us to deduce the relative branching ratio of the
two decay routes in question. 
With $c$ being adjusted appropriately we get a quantitative description of the
data both for $T_p = 750$ MeV \cite{bro02} and 775 MeV (see Figs. 3-5) with the
exception of the distributions for the invariant masses $M_{p\pi^+},
M_{p\pi^-}$ and $\widehat{\Theta}^{\pi^+\pi^-}_{\pi^+}$ (dashed lines in
Fig. 5). As shown in Ref.~\cite{bro02} ansatz (2) is able to heal this
deficiency (solid lines in Fig. 5) without destroying  the good agreement in
the other observables. We admit, however, that eq. (2) having the $\Delta$
propagator multiplying also the constant term, is a purely phenomenological ad
hoc ansatz. Apparently it is successful, but its physics contents  is not
(yet) fully 
understood. As suggested in Ref.~\cite{bro02} it possibly accounts effectively
for some final state interaction effect. Whereas in ansatz (1) the parameter
$c$ is defined as energy-independent, since all dynamics is taken into
account explicitly, the situation is not so clear with our phenomenological
ansatz (2); if some final-state interaction is absorbed here, an
energy dependence of $c'$ cannot be excluded a priori.

In Fig. 4 we compare calculations assuming different mixing scenarios for the
two $N^\ast$ decay routes to the data for those observables which are most
sensitive 
to the $N^\ast$ decays. Shown are the distributions of invariant masses
$M_{p\pi^+\pi^-}$~~and $M_{\pi^+\pi^-}$~~as well as~~$\delta_{\pi^+\pi^-}~=$
\newline  
$<\!\!\!)~({\bf k}_1,{\bf k}_2)$, i.e. the opening angle between the two
pion momenta in the overall cms. The latter distribution directly reflects the
squared decay amplitudes (1) and (2), respectively, averaged over all possible
moduli of pion momenta at given 
$\delta_{\pi^+\pi^-}$, i.e., $\sigma(\delta_{\pi^+\pi^-}) \sim (1 + b \cos
\delta_{\pi^+\pi^-})^2$ with the mixing coefficient $b$.

In case of eq. (2) we have $b = c' < k_1k_2>$ where the brackets denote the
average over all possible combinations. For $b \ll 1$ the distribution
$\sigma(\delta_{\pi^+\pi^-})$ is essentially linear in 
$b$, whereas this dependence gets quadratic for $b \gg 1$. Shown in Fig. 4 are
calculations for pure phase space and for transitions via either the $N^\ast
\to N\sigma$ route $(b = 0)$ or the $N^\ast \to \Delta \pi \to N\sigma$ route.
In order to illustrate the sensitivity of the data to the
mixing of both routes, calculations are also shown with $c' = -37$ and 
$-61\left(\mbox{GeV}/c\right)^{-2}$ corresponding to $b = -0.20$
and $-0.33$, respectively, at $T_p = 750$ MeV. The negative sign of the
coefficients reflects the destructive interference between both terms, which
is required by the data. If we fit $b$ for best reproduction 
of the data we obtain $b = -0.27(2)$ for $T_p = 750$ MeV and $b = -0.32(1)$
for $T_p = 775$ MeV, or   $c' = -50(4)$
and $-53(3) 
\left(\mbox{GeV}/c\right)^{-2}$, respectively. Both values
agree within their uncertainties, as they should if  $c'$ is energy
independent.  I.e. we not only observe the proper dependence in the angle
$<\!\!\!) ({\bf k}_1,{\bf k}_2)$, but also in the energy implied by 
$k_1k_2$ as the beam energy is changed, and with it the energy of
the $N^\ast$ excitation: for $T_p = 750$ MeV we
have $< M_{N^\ast} > = 1264$ MeV and for $T_p = 775$ MeV the average value is
$< M_{N^\ast} > = 1272$ MeV. 

\begin{figure}
\resizebox{0.45\textwidth}{!}{%
\includegraphics{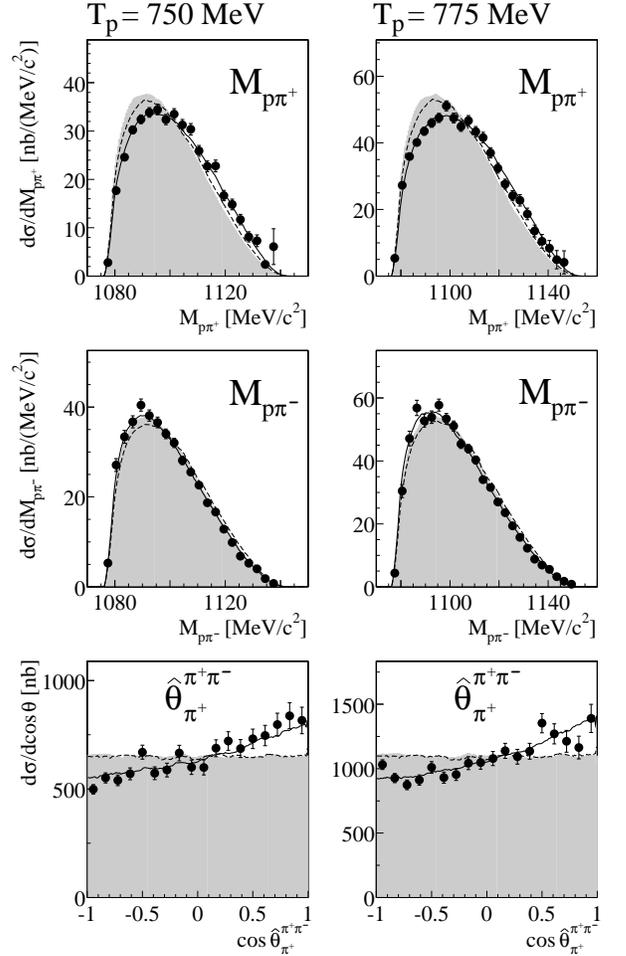}
}
\caption{Invariant masses of $p\pi^+$ and $p\pi^-$ systems (top and middle) as
  well as scattering angle of the $\pi^+$ in the $\pi\pi$-system  with respect
  to the summed pion momenta (bottom). The left side shows the results for
  750 MeV and the right one for 775 MeV. The shaded areas give phase space
  distributions, whereas dashed and solid lines show MC simulations according
  to ansatz (1) and (2), respectively.}
\end{figure}

Alternatively, if we use eq. (1) for the description of the data we arrive at
$c = 1.88(8)$ and $1.96(8) \left(\mbox{GeV}/c\right)^{-1}$ for $T_p
= 750$ and 775 MeV, respectively. 

Having fitted the parameters $c$ and $c'$, respectively, we can determine
the ratio of the partial decay widths for the routes $N^\ast \to \Delta \pi
\to N\pi\pi$ and $N^\ast \to N\sigma$ in dependence of the excited $N^\ast$
mass by
$$\begin{array}{l}
R(M_{N^\ast})  :=  {\displaystyle \frac{\Gamma_{N^\ast \to \Delta \pi \to
      N_{\pi\pi}}(M_{N^\ast})}{\Gamma_{N^\ast \to N\sigma}(M_{N^\ast})}} \\ 
~~~~~~~~~ \\
= {\displaystyle \frac{9}{8}}~c^2~(\mbox{or}~c'^2)
{\displaystyle \frac{\int \vert {\cal M}_{\Delta\pi} \vert^2 d M^2_{p\pi^+} d
    M^2_{\pi^+\pi^-}}{\int \vert {\cal M}_{N\sigma} \vert^2 d M^2_{p\pi^+} d
    M^2_{\pi^+\pi^-}}} \end{array}  \eqno(3)$$
with the matrix elements ${\cal M}_{N\sigma} = 1$ and ${\cal M}_{\Delta\pi} =
{\bf k}_1 \cdot {\bf k}_2 (3 D_{\Delta^{++}} + D_{\Delta^0})$ in case of
eq. (1). In case of eq. (2) these matrix elements are ${\cal M}_{N\sigma} = 
D_{\Delta^{++}}$ and ${\cal M}_{\Delta\pi} = {\bf k}_1 \cdot {\bf k}_2
D_{\Delta^{++}}$. Note that the integral is 
just the integration of the matrix element squared over the Dalitz plot in
dependence of the invariant masses $M^2_{p\pi^+}$ and $M^2_{\pi^+\pi^-}$. The
factor $9/8$ in eq. (3) is determined by isospin coupling coefficients and
accounts for the decay into channels other than $p(\pi^+\pi^-)_{I=l=0}$. If we
neglect spinflip contributions, then $2/3$ of both the $p^\ast \to p\sigma$
decay and of the $p^\ast \to \Delta \pi$ decay end up in the $p\pi^+\pi^-$
channel, i.e. the correction factor is unity instead of $9/8$

The results of these calculations are given in Table 1. For $T_p = 750$ MeV
and $T_p = 775$ MeV both equations lead to ratios $R(M_{N^\ast})$ which agree
within uncertainties. In this low-energy tail of the Roper resonance the ratio 
turns out to be very small, and is strongly energy-dependent as expected from
the ${\bf k}_1 \cdot {\bf k}_2$ dependence of $M_{\Delta\pi}$. 
We find $R(1272) \approx 1.5 \ast R(1264)$, i.e. a 50\% 
relative increase in the $N^\ast \to \Delta\pi$ route at the higher
energy. This increase is essentially due to the increase of $< k_1 k_2 >^2$,
which increases by more than 40\% by the 25 MeV increase in the beam
energy. The additional energy dependence in the $N^\ast \to \Delta \pi$ route
due to the $\Delta$ propagator in ansatz (1) is still of minor importance
here. In case of ansatz (2) such a small energy dependence could be
compensated easily by a small variation in the parameter $c'$ within its
statistical uncertainty. Hence with regard to this point the two data sets at
750 and 775 MeV are not yet able to discriminate between both equations.

\begin{table}
\caption{Ratio of the branching ratios for the decays $N^\ast \to \Delta\pi
  \to N\pi\pi$ and $N^\ast \to N\sigma$ in dependence of the excited $N^\ast$
  mass using ansatz (1) and (2), respectively, for the analysis of the data
  at $T_p = 750$ and 775 MeV. The extracted parameters $c$ and $c'$ are given
  in units of $\left(\mbox{GeV}/c\right)^{-1}$ and
  $\left(\mbox{GeV}/c\right)^{-2}$, respectively.}
\begin{tabular}{cccc}
& eq. (1) & & eq. (2) \\
\hline
& & & \\
$c$ or $c'$~~~~~~~ &1.88(8) & &--50(4) \\
($T_p = 750$ MeV)~~~~~~~ & & & \\
& & & \\
$c$ or $c'$~~~~~~~ &1.96(9) & &--53(3) \\
($T_p = 775$ MeV)~~~~~~~ & & & \\
& & & \\
R(1264)~~~~~~~ &0.034(4) & &0.030(5) \\
($T_p = 750$ MeV)~~~~~~~ & & & \\
& &  & \\
R(1272)~~~~~~~ &0.054(6) & &0.047(5) \\
($T_p = 775$ MeV)~~~~~~~& & & \\
& & & \\
R(1371)~~~~~~~ &1.0(1) & &0.6(1) \\
extrapolated~~~~~~~ & & & \\
& & & \\
R(1440)~~~~~~~&3.4(3) & &1.1(2) \\
extrapolated~~~~~~~ & & & \\
& & & \\
R(1440)~~~~~~~ & &4(2) & \\
PDG [1]~~~~~~~ & & & 
\end{tabular}
\end{table}

\vspace*{1cm}

However, the different appearance of the $\Delta$ propagator in eqs.
(1) and (2) will get discriminative if we go to higher energies. To
demonstrate this we extrapolate $R(M_{N^\ast})$ to the nominal resonance pole
at 1440 MeV/c$^2$ assuming eqs. (1) and (2) to hold also at higher energies
and taking for $c$ and $c'$ the average of the values obtained at $T_p = 750$
and 775 MeV. The different appearance of the $\Delta$ propagators in (1) and
(2) now leads to very different values.
In the case of the conventional ansatz, eq. (1), we get $R(1440) = 3.9(3)$
which 
is well within the range of the PDG  values of 4(2) \cite{pdg00}. In case of
eq. (2) the difference in the energy dependence of the two decay routes is
much smaller and we obtain only $R(1440) = 1.3(2)$. We note in passing that
due to 
the strong energy dependence of this ratio also the appropriate pole position
is crucial. If instead of the nominal Breit-Wigner mass pole position we use
the pole position evaluated from the speed plot of $\pi N$ phase shifts,
namely $M_{N^\ast} = 1371$ MeV/c$^2$ \cite{pdg00,kre00}, then the values for
the ratio decrease to 1.2(1) and 0.7(1), respectively, using eqs. (1) and
(2). 

In summary, the new set of differential data for the $pp \to pp\pi^+\pi^-$
reaction at $T_p = 775$ MeV supports the conclusion that this reaction is
dominated by the excitation of the Roper resonance and its decay into the
$N\pi\pi$ channels as derived recently \cite{bro02}
 from the analysis of the first
exclusive measurement at $T_p = 750$ MeV. The new data set gives a first
experimental evidence for a different energy dependence of the decay routes
$N^\ast \to \Delta \pi$ and $N^\ast \to N\sigma$. The decay branching of
$N^\ast \to \Delta\pi$ increases by 50\% relative to that of $N^\ast \to
N\sigma$, when increasing the incident proton energy from $T_p = 750$ MeV to
775 MeV, or equivalently when increasing the effective $N\pi\pi$ mass from
$M_{N^\ast} 
= 1264$ to $M_{N^\ast} = 1272$. In this very low-energy tail of the Roper
resonance  we find the $N^\ast \to N\sigma$ decay to be clearly dominant with
$R(1264) \cong 0.04$ and $R(1272) \cong 0.06$. These results are independent
of the 
ansatz used for the reaction amplitude. Though we observe the low-energy
region to be represented very well by a $N\sigma$ partition --- as
suggested e.g. in ref. \cite{kre00} --- we also see a small but rapidly
increasing influence of the $\Delta \pi$ partition.
Due to its $k_1 \cdot k_2$ dependence the $N^\ast \to \Delta\pi$ route is even
likely to finally take over at
higher energies. Extrapolating to the resonance pole we obtain $R(1440) =
3.4(3)$ and 1.1(2) using ansatz (1) and (2), respectively. Clearly this
extrapolation is strongly model-dependent. However, as we have demonstrated,
the $pp \to NN\pi\pi$ reaction offers the opportunity to experimentally
map out the energy dependence of the $N^\ast \to N\pi\pi$ decay systematically
up to the resonance pole by successively increasing the incident proton energy
--- a program which is currently pursued at CELSIUS-WASA. 

We acknowledge the continuous help of the TSL/ISV personnel and the support
by DFG (European Graduate School 683) and BMBF (06 TU 987).

\end{document}